\documentclass[11pt]{article}
\usepackage{a4wide}
\usepackage{bm} 
\usepackage{amssymb}
\usepackage{graphicx}
\usepackage{subfigure}
\usepackage{epsfig}
\usepackage[tbtags]{amsmath}
\usepackage{exscale}

\begin{document}

\setcounter{footnote}{0}

\newcommand{\lp}{\ell_{\mathrm P}}

\newcommand{\md}{{\mathrm{d}}}
\newcommand{\tr}{\mathop{\mathrm{tr}}}
\newcommand{\sgn}{\mathop{\mathrm{sgn}}}

\newcommand*{\R}{{\mathbb R}}
\newcommand*{\N}{{\mathbb N}}
\newcommand*{\Z}{{\mathbb Z}}
\newcommand*{\Q}{{\mathbb Q}}
\newcommand*{\C}{{\mathbb C}}

\newcommand{\gtt}{g_{\theta\theta}}
\newcommand{\gxx}{g_{xx}}
\newcommand{\bgtt}{\bar{g}_{\theta\theta}}
\newcommand{\bgxx}{\bar{g}_{xx}}
\newcommand{\ptt}{\pi^{\theta\theta}}
\newcommand{\pxx}{\pi^{xx}}
\newcommand{\bptt}{\bar{\pi}^{\theta\theta}}
\newcommand{\bpxx}{\bar{\pi}^{xx}}
\newcommand{\tptt}{\tilde{\pi}^{\theta\theta}}
\newcommand{\tpxx}{\tilde{\pi}^{xx}}
\newcommand{\kp}{\ensuremath{K_\varphi}}
\newcommand{\kx}{\ensuremath{K_x}}
\newcommand{\nx}{N^{x}}
\newcommand{\ex}{\ensuremath{E^x}}
\newcommand{\ep}{\ensuremath{E^\varphi}}
\newcommand{\gttd}{\dot{g}_{\theta\theta}}
\newcommand{\gxxd}{\dot{g}_{xx}}
\newcommand{\epd}{\dot{E}^{\varphi}}
\newcommand{\exd}{\dot{E}^{x}}
\newcommand{\gp}{\ensuremath{\Gamma_\varphi}}
\newcommand{\mr}{\ensuremath{\sqrt{\frac{2M}{x}}}}
\newcommand{\mrl}{\ensuremath{\sqrt{2Mx}}}
\newcommand{\be}{\begin{equation}}
\newcommand{\ee}{\end{equation}}
\newcommand{\bea}{\begin{eqnarray}}
\newcommand{\eea}{\end{eqnarray}}
\newcommand{\dif}{\mathrm{d}}
\newcommand{\ag}{a_{\gamma}}
\newcommand{\kpb}{\ensuremath{\bar{K}_\varphi}}
\newcommand{\kxb}{\ensuremath{\bar{K}_x}}
\newcommand{\exb}{\ensuremath{\bar{E}^x}}
\newcommand{\epb}{\ensuremath{\bar{E}^\varphi}}
\newcommand{\gpb}{\ensuremath{\bar{\Gamma}_\varphi}}
\newcommand{\ab}{\bar{\alpha}}
\newcommand{\aone}{\alpha_{1}}
\newcommand{\at}{\alpha_{2}}
\newcommand{\ath}{\alpha_{3}}
\newcommand{\afo}{\alpha_{4}}
\newcommand{\af}{\alpha_{5}}
\newcommand{\exbp}{\bar{E}^{x'}}
\newcommand{\vt}{\vartheta}
\newcommand{\vp}{\varphi}

\begin{center}
{\Large New second derivative theories of gravity for spherically symmetric spacetimes} \\
\vspace{1.5em}
Rakesh Tibrewala \footnote{e-mail address: {\tt rtibs@cts.iisc.ernet.in}}
\\
\vspace{1em}
Indian Institute of Science Education and Research,
CET Campus, Trivandrum 695016, India \footnote{Present address: Centre for High Energy Physics, Indian Institute of Science, Bangalore}
\end{center}

\begin{abstract}
We present new second derivative, generally covariant theories of gravity for spherically symmetric spacetimes (general covariance is in the $t-r$ plane) belonging to the class where the spherically symmetric Einstein-Hilbert theory is modified by the presence of $\gtt$ dependent functions. In $3+1$ dimensional vacuum spacetimes there is three-fold infinity of freedom in constructing such theories as revealed by the presence of three arbitrary $\gtt$ dependent functions in the Hamiltonian (matter Hamiltonian also has the corresponding freedom). This result is \emph{not} a contradiction to the theorem of Hojman et. al. \cite{hojman} which is applicable to the full theory whereas the above conclusion is for symmetry reduced sector of the theory (which has a much reduced phase space). In the full theory where there are no special symmetries, the result of Hojman et. al. will continue to hold. In the process we also show that theories where the constraint algebra is deformed by the presence of $\gtt$ dependent functions - as is the case in the presence of inverse triad corrections in loop quantum gravity - can always be brought to the form where they obey the standard (undeformed) constraint algebra by performing a suitable canonical transformation. We prove that theories obtained after performing canonical transformation are inequivalent to the symmetry reduced general relativity and that the resulting theories fall within the purview of the theories mentioned above.  

\end{abstract}


\section{Introduction}

Symmetry reduced models play a very important role in theories of gravity, both classical and quantum. The nonlinear nature of Einstein field equations or the complicated nature of the Hamiltonian constraint in the canonical formulation of the theory makes it prohibitively difficult to solve the classical or the quantum theory in generality. The best one can do is to take recourse in symmetry reduced models - the so called mini-superspace models (homogeneous cosmological models with only finitely many degrees of freedom) and the midi-superspace models (models with high degree of symmetry which nevertheless have infinite degrees of freedom like the spherically symmetric models) in addition to considering perturbations around these exact symmetry reduced models.

The usefulness of these models is obvious as the high degree of symmetry allows one to obtain exact solutions which help build intuition about the full theory and also allows one to test the general conclusions drawn on the basis of the full theory in simpler contexts. Interestingly, apart from giving considerable analytical control, it turns out that these models have a lot of physical relevance as is revealed by the successes of the homogeneous Friedmann-Robertson-Walker (FRW) cosmology, an example of mini-superspace model, as well as by the classical and semi-classical successes of black hole physics which often uses spherically symmetric midi-superspace models.

However, the requirement of general covariance of the theory puts stringent constraints on the form the theory can take. What this means is that one is stuck with a single theory to work with, to wit, Einstein's general relativity \cite{hojman} (unless one is willing to include higher derivative terms in the form of higher powers of curvature tensor and its contractions \cite{lovelock}). This then implies that in the absence of higher curvature terms the symmetry reduced models derived thereof are also unique.

This is in contrast to non-gravitational field theories where even after imposing Lorentz invariance (and the gauge principle where gauge fields are present), sufficient freedom still remains to construct more than one model (obtained by changing the potential function, for instance). The theoretical advantages of having more than one model available are too obvious to be elaborated upon. 

In this paper we will partly overcome this deficiency of gravitational theory by showing that as far as spherically symmetric midi-superspace models are concerned, one can have many generally covariant theories (general covariance restricted to the $t-r$ plane on account of spherical symmetry) without involving higher derivative/higher curvature corrections. That such a possibility exists was, in a sense, already noticed in \cite{modifiedhorizon, MartinPaily} in the context of certain loop quantum gravity (LQG) inspired corrections (see section 5 below) though its meaning was not fully appreciated.

Explicitly, restricting to $3+1$ dimensions and using the Hamiltonian approach, we will find that there is three-fold infinity of freedom in the construction of generally covariant theories for spherically symmetric spacetimes in the absence of matter degrees of freedom, the three-fold infinity of freedom appearing in the form of three arbitrary functions of the metric component $\gtt$ in the Hamiltonian (equivalently, the Lagrangian) of the theory (matter Hamiltonian also has the corresponding freedom) .

To obtain these new theories we will be using the criteria of Hojman et. al. \cite{hojman} that the constraint algebra obeyed by the constraints in the Hamiltonian formulation of a generally covariant theory should be identical to the hypersurface deformation algebra (which encodes the kinematics of deformation of an hypersurface embedded in spacetime).

In fact, the objective of the present work can be motivated differently by focussing on the results in \cite{hojman} according to which the canonical representation of the generators of the deformations of a hypersurface (embedded in a $3+1$ dimensional Riemannian spacetime) on the phase space with the intrinsic metric of the hypersurface and its conjugate momentum as the sole canonical variables is unique (when there are no symmetries imposed on the spacetime) and is precisely the one that follows from the Einstein-Hilbert action. 

The only freedom available in constructing the phase space representation of these generators -- super-Hamiltonian (for deformations normal to the hypersurface) and super-momentum (for deformations tangential to the hypersurface) -- is a canonical transformation affecting only the momenta conjugate to the intrinsic metric of the hypersurface with the new momenta differing from the old by a factor dependent only on the intrinsic geometry of the hypersurface.

In \cite{hojman} the embedding spacetime is generic with no special symmetries so that the number of generators is the same as the dimensionality of the spacetime. Since classically symmetry reduction is exact, one might expect that the operations of (i) constructing the unique representation of generators of hypersurface deformation as per \cite{hojman} and (ii) imposing some symmetry on the spacetime (spherical symmetry, for instance) should commute and one should end up with identical results. In other words, it might seem that whether we construct representation of the surface deformation algebra in the full theory first and then impose some spacetime symmetry on this representation or we first impose spacetime symmetry and then construct the representation of the symmetry reduced theory the result would remain the same.

However, as will be shown in this paper, this is not the case and imposition of (spherical) symmetry on spacetime before constructing the representation(s) of generators leads to several inequivalent representations. Specifically, we will find that the generator for normal deformations, the super-Hamiltonian, is not unique (the Hamiltonian of spherically reduced general relativity is just one among infinity of inequivalent Hamiltonians possible for these spacetimes). We will prove that the modified (symmetry reduced) theory is \emph{inequivalent} to the (symmetry reduced) general relativity by showing that it is not possible to construct a canonical transformation, depending only on the three geometry \cite{hojman}, which leads from the new representation(s) to the general relativistic representation. 

Here we would like to emphasize that the result of this paper is \emph{not} in contradiction with those of Hojman et. al. \cite{hojman}. The proof in \cite{hojman} is for full general relativity (without assuming
any underlying symmetry) and therefore the corresponding phase space is that of the full theory ($12\infty^{3}$
dimensional before imposing constraints) whereas here we are looking at the (spherical) symmetry
reduced sector of the theory and therefore the phase space is much smaller (only $4\infty$ dimensional
before imposing constraints). The uniqueness result of Hojman et. al. need not apply in
such a scenario. In the full theory, where there are no special spacetime symmetries, the result in \cite{hojman} will, obviously, continue to hold.

In the process of deriving these new theories we will also end up showing that for spherically symmetric spacetimes even when the constraint algebra is deformed by the presence of $\gtt$ dependent corrections (due to the presence of $\gtt$ dependent modifications of the Hamiltonian), as is essentially the case in LQG in the presence of the so called \emph{inverse triad corrections}, it is always possible to perform a canonical transformation so that the constraint algebra becomes undeformed (and thus has the structure of the standard surface deformation algebra) such that the resulting theory falls within the realm of the theories that will be constructed in this paper.

This was, to a certain extent, already done in \cite{canonical transformation} but only for a particular case of inverse triad corrections. Here we generalize the construction to generic $\gtt$ dependent modifications in the Hamiltonian (thus showing that thinking in terms of the inverse triad corrections of LQG is not necessary). In addition, we prove that the theories thus obtained are inequivalent to symmetry reduced version of general relativity (something which was not shown in \cite{canonical transformation}). Additionally, the aim is to highlight that although the physical motivation for the new theories might lie in quantum gravity (since we expect that quantization of spherically symmetric models should lead to certain modifications without spoiling the constraint algebra which encodes the diffeomorphism invariance of the theory), there is nothing intrinsically `quantum' about the new Hamiltonians.

These new theories and their solutions can serve as useful toy models for testing whether the conclusions based on  general relativity apply to a generic generally covariant theory or not and can serve as useful testing ground for quantum field theory on curved backgrounds as also for theories of quantum gravity, all of which frequently use the spherical symmetry ansatz. In fact, in the context of quantum gravity, these new theories will also provide an opportunity to test the reliability of the conclusions based on midi-superspace quantization.

In short, the aim of this work is to show that in the context of spherically symmetric spacetimes (i) a deformed constraint algebra that results in the presence of inverse triad corrections in LQG (or more generally, a deformed constraint algebra resulting from a $\gtt$ dependent modifications of the Hamiltonian constraint) can always be made to have the standard (or classical) form (as in \cite{canonical transformation} where this was done for a special case), (ii) this procedure gives new second derivative theories of spherically symmetric gravity which are inequivalent to the spherical reduction of general relativity and (iii) even though the construction is motivated by studies in a quantum theory of gravity, specifically LQG, the new theories are entirely classical thus providing several second derivative models to work with.

In the next section we give a brief outline of the proof of \cite{hojman} identifying the key assumptions of the proof and highlighting the possible freedom available in the representation of the generators of surface deformation. Section 3 briefly discusses the standard canonical formulation of spherically symmetric general relativity along with showing the explicit structure of the surface deformation algebra. In section 4 we present a detailed derivation of the new theories (without emphasizing LQG motivation much since we want to emphasize the classical aspects of the construction; a discussion in the light of LQG is relegated to section 6). In section 5 we give a proof that the new theories are inequivalent to the (spherical) symmetry reduced version of general relativity by showing that the difference in the canonical momenta of the new and the old theories is not just dependent on the three geometry but also on how the three geometry is embedded in the four dimensional spacetime. This implies that the new and the old theories are not related by a canonical transformation. Here we also show that inclusion of matter allows for extra freedom in the form of additional $\gtt$ dependent function(s). In section 7 we present the possible significance of these new theories and additionally discuss the limitations due to the imposition of spherical symmetry and possible ways to go beyond spherical symmetry. We conclude in section 8. 

\section{Summary of the proof of Hojman et. al.}
The aim of the work of Hojman et. al. \cite{hojman} is to obtain the canonical formulation of the general theory of relativity from plausible first principles, without going through the usual route of starting with the Einstein-Hilbert action. In the usual formulation, one starts with the Arnowitt-Deser-Misner (ADM) metric \cite{adm} obtained by splitting the spacetime metric $^{4}g_{\mu\nu}$ into the spatial metric $g_{ij}$ on space-like hypersurfaces of constant $t$, the lapse function $N$ and the shift vector $N^{i}$ (Greek indices $(\mu, \nu...)$ take values $(0,1,2,3)$ and Latin indices $(i, j...)$ take values $(1,2,3)$)
\be \label{adm metric}
\md s^{2}=-N^{2}dt^{2}+g_{ij}(dx^{i}+N^{i}dt)(dx^{j}+N^{j}dt).
\ee

To obtain the canonical formulation, the Einstein-Hilbert action is expressed in terms of the quantities defined on the three dimensional spatial slice and takes the form
\be \label{einstein hilbert action in adm formulation}
S=\int\md t\, L=\frac{1}{16\pi G}\int\md t\int\md^{3}x N\sqrt{g}(^{(3)}R+K_{ij}K^{ij}-K^{2}).
\ee 
In the above equation $G$ is Newton's constant, $g$ is the determinant of the spatial metric, $^{(3)}R$ is the Ricci scalar for $g_{ij}$, $K_{ij}$ is the extrinsic curvature of the hypersurface and $K=g^{ij}K_{ij}$ (and we take the cosmological constant to be zero). 

The Hamiltonian of the theory is obtained by performing the Legendre transformation noting that since time derivatives of the lapse and the shift functions do not appear, they do not contribute to the Legendre transformation. Introducing the canonical momentum $\pi^{ij}$ conjugate to $g_{ij}$  and going through the standard procedure one finds that the action \eqref{einstein hilbert action in adm formulation} (for pure geometrodynamics) takes the form
\be \label{adm action}
S=\int\md t\int\md^{3}x(\pi^{ij}g_{ij,0}-N\mathcal{H}-N^{i}\mathcal{D}_{i}),
\ee
where $\mathcal{H}\equiv \mathcal{H}(g_{ij},\pi^{ij})$ is the super-Hamiltonian and $\mathcal{D}_{k}\equiv \mathcal{D}_{k}(g_{ij},\pi^{ij})$ is the super-momentum. 

This action when varied with respect to the canonical variables $(g_{ij}$, $\pi^{ij})$ leads to Hamilton's equations and when varied with respect to the lapse function $N$ and the shift vector $N^{i}$ gives, respectively, the so called Hamiltonian constraint $\mathcal{H}=0$ and the diffeomorphism constraint $\mathcal{D}_{i}=0$. For consistency of the overall formulation, the constraints must be preserved from one slice to the next implying that they form a closed system under Poisson bracket. 

As discussed in \cite{hojman}, this algebra turns out to have a definite structure and in fact mimics the algebra of the generators of deformations of a hypersurface embedded in a Riemannian spacetime (the so called hypersurface deformation algebra). If $\mathcal{H}(x)$ denotes the generator of deformations normal to the hypersurface and $\mathcal{D}_{i}(x)$ the corresponding generators for tangential deformations (to be distinguished from $\mathcal{H}(g_{ij},\pi^{ij})$ and $\mathcal{D}_{i}(g_{ij},\pi^{ij})$ defined earlier) then we have
\bea \label{surface deformation algebra hh}
\{\mathcal{H}(x), \mathcal{H}(\tilde{x})\} &=& \mathcal{D}^{i}(x)\delta_{,i}(x,\tilde{x})-\mathcal{D}^{i}(\tilde{x})\delta_{,i}(\tilde{x},x), \\
\label{surface deformation algebra dh}
\{\mathcal{D}_{i}(x), \mathcal{H}(\tilde{x})\} &=& \mathcal{H}(x)\delta_{,i}(x,\tilde{x}), \\
\label{surface deformation algebra dd}
\{\mathcal{D}_{i}(x), \mathcal{D}_{j}(\tilde{x})\} &=& \mathcal{D}_{i}(\tilde{x})\delta_{,j}(x,\tilde{x})+\mathcal{D}_{j}(x)\delta_{,i}(x,\tilde{x}).
\eea

Note that the spatial metric $g_{ij}$ explicitly enters relation \eqref{surface deformation algebra hh} in raising the index on $\mathcal{D}_{i}$. Hojman et. al. \cite{hojman} use the structure of this algebra as the basic principle on which to base the canonical theory. That is, instead of taking the Einstein-Hilbert action as given and arriving at the super-Hamiltonian and the super-momentum obeying this structure, the algebra in \eqref{surface deformation algebra hh}-\eqref{surface deformation algebra dd} is taken as the starting point and one asks for the representation of the generators of the normal deformation $\mathcal{H}(x)$ and the tangential deformation $\mathcal{D}_{i}(x)$ on the geometrodynamical phase space $(g_{ij}, \pi^{ij})$.

The principle of path independence leads to the conclusion that these generators must obey the constraints $\mathcal{H}=0$ and $\mathcal{D}_{i}=0$. To construct the representation of the generator of the tangential deformations $\mathcal{D}_{i}$, of the three closing relations above, only relation \eqref{surface deformation algebra dd} is made use of. One first calculates the change in a dynamical variable $F$ under a tangential deformation in two different ways - (i) using the relation $\delta F=\mathcal{L}_{\delta N^{i}}F$ and (ii) using the evolution equation 
\be \label{dynamical evolution}
\delta F=\int\md x\{F, \mathcal{H}(x)\}\delta N(x)+\int\md x\{F, \mathcal{D}_{i}(x)\}\delta N^{i}(x),
\ee
with $\delta N=0$. On equating these one can find the functional derivatives of $\mathcal{D}_{i}$ with respect to the three metric $g_{ij}$ as also with respect to the conjugate momentum $\pi^{ij}$. Use of certain integrability condition along with the closing relation \eqref{surface deformation algebra dd} then fixes the form of the super-momentum. 

To construct the representation of the super-Hamiltonian $\mathcal{H}$ one first uses the closing relation \eqref{surface deformation algebra dh} to find that the super-Hamiltonian must be a scalar density of weight one. Using \eqref{dynamical evolution} to find the change in the metric $g_{ij}$ under normal deformation $(\delta N^{i}=0)$ then shows that $\mathcal{H}$ should be a local function of the conjugate momenta $\pi^{ij}$. Next, assuming the argument of time-irreversibility it is concluded that $\mathcal{H}$ should be an even functional of  momenta $\pi^{ij}$ (in \cite{kuchar geometrodynamics lagrangian} it is shown that this assumption is not necessary when using the Lagrangian approach to construct the representation of the surface deformation algebra). 

Finally, a series expansion of $\mathcal{H}$ is made in even powers of $\pi^{ij}$ and the coefficients of this expansion (which are functionals of the three metric) are determined term by term using the closing relation \eqref{surface deformation algebra hh}. It turns out that the super-Hamiltonian and the super-momentum have the same form as obtained from the canonical formulation of the Einstein-Hilbert theory in \eqref{adm action}. The only freedom available is a canonical transformation 
\be \label{freedom in definition of momenta as per hojman et al}
\pi^{ij}(x)\rightarrow\pi^{ij}(x)+\delta\Lambda/\delta g_{ij}(x)
\ee 
where $\Lambda$ is an arbitrary scalar functional of the metric (and so that it does not depend on the labelling, it is actually a functional of the three geometry $\Lambda\equiv\Lambda[^{3}\mathcal{G}]$).

\section{Canonical formulation of classical general relativity with spherical symmetry}

In this section we recapitulate the essentials of the canonical formulation of spherically symmetric general relativity limiting ourselves (for simplicity) to vacuum spacetimes. This is followed by presenting the structure of the constraint algebra for these spacetimes.

For spherically symmetric spacetimes the ADM metric in \eqref{adm metric} is
\be \label{classical metric usual variables}
\md s^{2}=-N^{2}\md t^{2}+\gxx\left(\md x+\nx\md t\right)^{2}+\gtt\md\Omega^{2}.
\ee
Here $\gxx$, $\gtt$ are the only dynamical variables and because of spherical symmetry these are functions of the time coordinate $t$ and the radial coordinate $x$ only. For the same reason, $\nx$ is the only non-zero component of the shift vector. From \eqref{classical metric usual variables} we read-off the metric on the spatial slice $\Sigma$:
\be \label{spatial metric}
\md s^{2}|_{\Sigma}=\gxx\md x^{2}+\gtt\md\Omega^{2}.
\ee
Using this metric, the Lagrangian \eqref{einstein hilbert action in adm formulation} becomes
\bea \label{sph symm lagrangian}
L&=&\int\md x\bigg[-\frac{(\gttd)^{2}\sqrt{\gxx}}{8GN\gtt}-\frac{\gttd\gxxd}{4GN\sqrt{\gxx}}+\frac{\nx\gxx'\gttd}{4GN\sqrt{\gxx}}+\frac{\nx\gtt'\gxxd}{4GN\sqrt{\gxx}}+\frac{\nx\gtt'\gttd\sqrt{\gxx}}{4GN\gtt} \nonumber \\
&&-\frac{\nx N^{x'}\gtt'\sqrt{\gxx}}{2GN} +\frac{N^{x'}\gttd\sqrt{\gxx}}{2GN}-\frac{(\nx)^{2}\gtt'\gxx'}{4GN\sqrt{\gxx}}-\frac{(\nx)^{2}(\gtt')^{2}\sqrt{\gxx}}{8GN\gtt}+\frac{N(\gtt')^{2}}{8G\gtt\sqrt{\gxx}} \nonumber \\
&&+\frac{N\sqrt{\gxx}}{2G}-\frac{N\gtt''}{2G\sqrt{\gxx}}+\frac{N\gtt'\gxx'}{4G(\gxx)^{3/2}}\bigg],
\eea
where a dot represents derivative with respect to $t$ while a prime denotes derivative with respect to $x$ (and the angular coordinates have been integrated over).

Using $\pi^{ij}(x)=\delta L/\delta g_{ij}(x)$ we find the canonical momentum conjugate to $\gxx$ and $\gtt$:
\bea \label{classical momenta pxx}
\pxx &=& \frac{-\dot{g}_{\theta\theta}+\nx\gtt'}{4GN\sqrt{\gxx}}, \\
\label{classical momenta ptt}
\ptt &=& \frac{1}{4GN}\bigg(-\frac{\dot{g}_{\theta\theta}\sqrt{\gxx}}{\gtt}-\frac{\dot{g}_{xx}}{\sqrt{\gxx}}+\frac{\nx\gtt'\sqrt{\gxx}}{\gtt}+\frac{\nx\gxx'}{\sqrt{\gxx}}+2N^{x'}\sqrt{\gxx}\bigg).
\eea
The pair $(\gtt, \ptt)$ and $(\gxx, \pxx)$ coordinatizes the phase space of the theory and obey the Poisson bracket relations
\be \label{classical poisson bracket}
\{\gtt(x), \ptt(y)\}=\delta(x,y), \quad \{\gxx(x), \pxx(y)\}=\delta(x,y)
\ee
with all other Poisson brackets among these variables being identically equal to zero.

The (total) Hamiltonian of the theory,
\be \label{total hamiltonian general expression}
H_{T}=\int\md x[\pxx\dot{g}_{xx}+\ptt\dot{g}_{\theta\theta}]-L,
\ee
is
\bea \label{classical total hamiltonian}
H_{T} &=& \int\md\,x N\bigg[-4G\sqrt{\gxx}\pxx\ptt+\frac{2G\gxx^{3/2}(\pxx)^{2}}{\gtt}-\frac{\sqrt{\gxx}}{2G}-\frac{\gxx'\gtt'}{4G\gxx^{3/2}}-\frac{\gtt'^{2}}{8G\sqrt{\gxx}\gtt}+\frac{\gtt''}{2G\sqrt{\gxx}}\bigg] \nonumber \\
&& +\int\md x\,\nx\left[\gtt'\ptt-\gxx'\pxx-2\gxx\pi^{xx'}\right].
\eea
From this expression we can read-off the Hamiltonian constraint (in the integrated form)
\bea \label{classical hamiltonian constraint metric variables} 
H[N]&=&\int \md x\, N\bigg[-4G\sqrt{\gxx}\pxx\ptt+\frac{2G\gxx^{3/2}(\pxx)^{2}}{\gtt}-\frac{\sqrt{\gxx}}{2G}-\frac{\gxx'\gtt'}{4G\gxx^{3/2}}-\frac{\gtt'^{2}}{8G\sqrt{\gxx}\gtt} \nonumber \\
&&+\frac{\gtt''}{2G\sqrt{\gxx}}\bigg]\approx0
\eea
and the diffeomorphism constraint
\be \label{classical diffeomorphism constraint metric variables}
D[\nx]=\int \md x\,\nx\left[\gtt'\ptt-\gxx'\pxx-2\gxx\pi^{xx'}\right]\approx0.
\ee

Note that the classical super-Hamiltonian is quadratic in the canonical momenta and that the three metric appears with spatial derivatives of order at most two while the super-momentum is linear in conjugate momenta. 
These constraints obey the following Poisson bracket algebra
\bea \label{classical hh algebra}
\{H[N],H[M]\} &=& D[\gxx^{-1}(NM'-N'M)], \\
\label{classical dh algebra}
\{D[N^{x}],H[N]\} &=& H[N'N^{x}], \\
\label{classical dd algebra}
\{D[N^{x}],D[M^{x}]\} &=& D[N^{x}M^{x'}-N^{x'}M^{x}].
\eea

Comparing this algebra of constraints with the surface deformation algebra in equations \eqref{surface deformation algebra hh}-\eqref{surface deformation algebra dd} it is clear that the former is nothing but the integrated form of the latter (with spherical symmetry imposed) and provides the motivation for the derivation in \cite{hojman} of reversing the standard procedure and arriving at the Hamiltonian of general relativity by constructing the representation of the generators of surface deformation algebra).

\section{New second derivative theories of gravity for spherically symmetric spacetimes}

In this section we construct new generally covariant theories for spherically symmetric spacetimes involving no more than second derivative of the metric (general covariance limited to the $t-r$ plane because of spherical symmetry). The physics motivation (or the idea) for the construction of these theories has its origin in the study of certain LQG inspired corrections in the context of spherically symmetric models. According to our view point, however, this connection with quantum gravity (LQG or otherwise) is incidental and the new theories, if one wishes, can be regarded as completely classical in that the corresponding Hamiltonian (Lagrangian) need not involve the Planck constant (or the Planck length). Readers interested in the quantum gravity motivation should refer to section 6 below.

Instead of quantum gravity motivation one can start by simply asking the following question - from the result in \cite{hojman} we know that in a four dimensional spacetime without any underlying spacetime symmetry (in other words, in the absence of any Killing vectors) the representation of the surface deformation algebra on the geometrodynamic phase space is unique and is exactly what is obtained by starting with the Einstein-Hilbert action. Now suppose we impose an underlying spacetime symmetry on the four dimensional spacetime manifold (equivalently, introduce Killing vectors). Then we can ask whether for this symmetry reduced theory (and, correspondingly, reduced phase space) do we obtain a unique representation of the symmetry reduced surface deformation algebra or whether for spacetimes with Killing vectors more than one representations are possible?

The significance of this question can be motivated in yet another way. Since classically symmetry reduction is exact, one might expect that the operations of (i) constructing the unique representation of generators of hypersurface deformation as per \cite{hojman}, followed by (ii) imposing some symmetry on the spacetime (spherical symmetry, for instance) should commute. In other words, one might think that at least classically, irrespective of  whether we construct the representation of the surface deformation algebra in the full theory first and then impose some spacetime symmetry (spherical symmetry in the present case) or we first impose spacetime symmetry on the underlying manifold and then construct the representation of the symmetry reduced theory the result would remain the same.

As will be shown below, this is not the case and the operations of constructing the representation of the surface deformation algebra and of imposing spacetime symmetry do not commute. We will explicitly show that when we impose spherical symmetry on the underlying manifold first and then find the corresponding representation, we find that there are infinitely more representations possible (different representations of the constraints correspond to different theories). On the other hand, if we construct the representation of the full theory first (which is unique and equivalent to general relativity by \cite{hojman}) and then impose (spherical) symmetry, we would have obtained a unique representation - to wit, (spherical) symmetry reduced version of general relativity.

At the outset we would like to make it clear that this non-uniqueness of representation for symmetry reduced models is in no way in conflict with the result of Hojman et. al \cite{hojman}. The proof in \cite{hojman} is for full general relativity without assuming any underlying spacetime symmetry (no Killing vectors) and therefore the corresponding phase space is that of the full theory - 12$\infty^{3}$ dimensional before imposing (four) constraints. For spherically reduced sector, on the other hand, the phase space is much smaller - only 4$\infty$ dimensional before imposing (two) constraints. The uniqueness result of Hojman et. al. need not apply in such a scenario.

After this long prelude we are now ready to construct the new theories (or representations of the Hamiltonian constraint) for spherically symmetric spacetimes. Since we would be using the canonical approach we begin with the ADM metric \eqref{classical metric usual variables} but trade the variables $(\gxx,\gtt)$ for two new variables $(\ep,\ex)$ such that
\bea \label{relation between metric variables and lqg variables}
\ex(t,x) &=& \gtt(t,x), \nonumber \\
\ep(t,x) &=& \sqrt{\gtt\gxx},
\eea
so that the metric is
\be \label{classical metric lqg variables} 
{\rm d}s^2=-N^2{\rm d}t^2+\frac{\ep\,^2}{E^{x}}({\rm d}x+N^x{\rm
  d}t)^2+ E^{x}{\rm d}\Omega^2.
\ee
The use of these variables is motivated from LQG (see section 6). However, as mentioned earlier, we need not think in terms of quantum gravity and the above choice can simply be thought of as a change of variables which simplifies some of the mathematical steps (especially in the context of the canonical transformation to be discussed later). 

Momenta conjugate to $(\ep, \ex)$ are denoted $(\kp,\kx)$ respectively and obey the commutation relations
\be \label{classical poisson bracket lqg}
\{\kp(x), \ep(y)\}=G\delta(x,y), \, \{\kx(x), \ex(y)\}=2G\delta(x,y).
\ee
 
In terms of these variables and for vacuum, the classical Hamiltonian constraint $H[N]$ \eqref{classical hamiltonian constraint metric variables} and the diffeomorphism constraint $D[N^{x}]$ \eqref{classical diffeomorphism constraint metric variables} take the form:
\bea \label{classical hamiltonian constraint lqg variables}
H[N] &=& -\frac{1}{2G}\int \md x\, N\bigg[\frac{\kp^2\ep}{\sqrt{\ex}}+2\kp\kx\sqrt{\ex}+\frac{\ep}{\sqrt{\ex}} -\frac{(E^{x'})^{2}}{4\ep\sqrt{\ex}}-\frac{E^{x''}\sqrt{\ex}}{\ep} \nonumber \\
&&+\frac{E^{\varphi'}E^{x'}\sqrt{\ex}}{(\ep)^{2}} \bigg]\approx0, \\
\label{diffeomorphism constraint lqg variables}
D[N^x] &=& \frac{1}{2G}\int \md x\,N^x\bigg[2\kp'\ep-\kx E^{x'}\bigg]\approx0.
\eea

In LQG it turns out that inverse components of $\ex$ in the Hamiltonian do not have direct operator analog and lead to certain quantum corrections which, at a semi-classical level, are taken into account by replacing $1/(\ex)\rightarrow\alpha(\ex)/\ex$ (where $\alpha(\ex)$ is a quantum correction function derived using LQG techniques). We will have more to say about these corrections in section 6. For now, motivated by this observation, we consider a purely mathematical generalization whereby each term of the classical Hamiltonian \eqref{classical hamiltonian constraint lqg variables} is modified by an arbitrary $\ex$-dependent factor $\alpha_{i}(\ex)$. We emphasize that these $\alpha_{i}(\ex)$ are completely arbitrary and (for this section at least) they \emph{need not} be related to LQG or any other quantum theory of gravity. We thus consider the Hamiltonian  
\bea \label{generalized hamiltonian lqg variables}
\tilde{H}[N]&=&-\frac{1}{2G}\int \md x\, N\bigg[\frac{\aone\kp^2\ep}{\sqrt{\ex}}+2\at\kp\kx\sqrt{\ex}+\frac{\ath\ep}{\sqrt{\ex}}-\frac{\afo(E^{x'})^{2}}{4\ep\sqrt{\ex}}-\frac{\af E^{x''}\sqrt{\ex}}{\ep} \nonumber \\
&&+\frac{\alpha_{6}E^{\varphi'}E^{x'}\sqrt{\ex}}{(E^{\varphi})^{2}} \bigg],
\eea

In the canonical formulation the Hamiltonian constraint is the generator of the `time' translations whereas the diffeomorphism constraint only generates diffeomorphism within the spatial slice. Since it is the Hamiltonian constraint which generates the dynamics by inducing motion from one spatial slice to the next, a modified Hamiltonian will correspond to modified dynamics. However, we do not want to tinker with the spatial diffeomorphisms and for this reason we leave the diffeomorphism constraint \eqref{diffeomorphism constraint lqg variables} unmodified. In fact, as shown in \cite{hojman, kuchar geometrodynamics lagrangian} there is hardly any freedom in the construction of the diffeomorphism constraint and its form is essentially fixed by the requirement $\{F,D\}\delta\nx=\mathcal{L}_{\delta\overrightarrow{\nx}}F$ without requiring the use of \eqref{classical hh algebra}-\eqref{classical dd algebra}. 

We now calculate the Poisson bracket between two modified Hamiltonians \eqref{generalized hamiltonian lqg variables} to find
\bea \label{general anomalous algebra}
&&\{\tilde{H}[N], \tilde{H}[M]\}=\frac{1}{2G}\int\md x\, (NM'-N'M)\frac{\kp}{\ep}\bigg[(\at\afo-\aone\alpha_{6})E^{x'}+2(\at'\af-\at\af')\ex\bigg] \nonumber \\
&&+\frac{1}{2G}\int\md x\, (NM'-N'M)\frac{\ex}{(\ep)^{2}}\bigg[\at(2\af\kp'\ep-\alpha_{6}\kx E^{x'})+2\at(\af-\alpha_{6})\kp E^{\varphi'}\bigg].
\eea
We note that unlike the standard $H-H$ bracket \eqref{classical hh algebra}, the right side of the above equation is not simply related to the diffeomorphism constraint. Since we have left the diffeomorphism constraint unmodified, the bracket between two diffeomorphisms obviously does not change while the bracket involving $D-H$ is modified (whose exact form is not important for reasons that will become clear in the following). 

For the algebra to be anomaly-free the RHS of \eqref{general anomalous algebra} should depend only on the constraints. This will be so if
\be \label{anomaly freedom one}
\af=\alpha_{6},
\ee
and
\be \label{anomaly freedom two}
(\at\afo-\aone\alpha_{6})E^{x'}+2(\at'\af-\at\af')\ex=0.
\ee
Using these conditions it is found that the $D-H$ bracket retains its original form \eqref{classical dh algebra}.

Condition \eqref{anomaly freedom two} after using \eqref{anomaly freedom one} can be used to express one of the $\alpha$'s, say $\afo$, in terms of the others
\be \label{a four}
\afo=\frac{\aone\af}{\at}-2\left(\frac{\af}{\at}\frac{\md\at}{\md\ex}-\frac{\md\af}{\md\ex}\right)\ex.
\ee
Note that even after imposing these conditions we have four arbitrary functions $\aone, \at, \ath$ and $\af$ and the Poisson bracket between two Hamiltonians becomes
\be \label{deformed hh bracket}
\{\tilde{H}[N], \tilde{H}[M]\}=D[\at\af\ex(\ep)^{-2}(NM'-N'M)].
\ee
The algebra is thus deformed unless $\at\af=1$.  

To take care of the deformed algebra we will follow the procedure of \cite{canonical transformation} where the main point was to show that the deformed (or non-classical) constraint algebra for spherically symmetric spacetimes in the presence of $\ex$ dependent deformations is only an artifact of the particular choice for the phase space coordinates and that with a suitable canonical transformation the constraint algebra can be rendered classical. Actually that work focussed on the special case where all the $\alpha$'s in \eqref{generalized hamiltonian lqg variables} were identical $\alpha_{i}=\alpha\neq1, (i=1,2...6)$, so that the deformation factor was $\alpha^{2}$. Here we are generalizing that procedure to arbitrary $\alpha$ functions which would thus lead to the most general $\ex$ dependent modifications of the Hamiltonian which would nevertheless obey the standard surface deformation algebra and would thus correspond to most general spherically symmetric theory with such modifications. 

The canonical transformation is performed using the generating function $F_{3}=-\at\af\ex\kxb$ depending on the new coordinate $\kxb$ and the old momentum $\ex$ such that 
\bea \label{canonical transformation}
&& \exb=-\frac{\partial F_{3}}{\partial\kxb}=\at\af\ex, \\
&& \kx=-\frac{\partial F_{3}}{\partial\ex}=\left(\at\af+\af\ex\frac{\md\at}{\md\ex}+\at\ex\frac{\md\af}{\md\ex}\right)\kxb.
\eea
(As a side remark we note that had we continued working with the usual metric variables, it would have turned out that the canonical transformation involve both the pairs $(\gxx,\pxx)$ and $(\gtt,\ptt)$ unlike the present case where only the $(\ex,\kx)$ pair is involved and it is generally simpler to perform canonical tranformations involving only one canonical pair and justifies the choice of $(\ex,\ep)$ over $(\gtt,\gxx)$ for the analysis of this section.)
 
Interestingly, in terms of the transformed variables, the super-momentum retains its classical form (see \cite{canonical transformation} for details)
\be \label{super momentum after canonical transformation}
D[\nx]=\frac{1}{2G}\int\md x\,\nx[2\kp'\ep-\kxb\bar{E}^{x'}],
\ee 
while the Hamiltonian constraint takes the form
\bea \label{super hamiltonian after canonical transformation}
\bar{H}[N]&=&-\frac{1}{2G}\int\md x\,N\bigg[\frac{A_{1}\kp^{2}\ep}{\sqrt{\exb}}+2A_{2}\kp\kxb\sqrt{\exb}+\frac{A_{3}\ep}{\sqrt{\exb}}-\frac{A_{4}(\bar{E}^{x'})^{2}}{4\ep\sqrt{\exb}}-\frac{\bar{E}^{x''}\sqrt{\exb}}{A_{2}\ep} \nonumber \\
&&+\frac{E^{\varphi'}\exbp\sqrt{\exb}}{A_{2}(\ep)^{2}}\bigg].
\eea
In the above expression
\bea
A_{1}&=&\aone\sqrt{\at\af}\,, \\
A_{2}&=&\frac{\at^{5/2}\af^{3/2}}{\at\af-\exb\at\frac{\md\af}{\md\exb}-\exb\af\frac{\md\at}{\md\exb}}\,, \\
A_{3}&=&\ath\sqrt{\at\af}\,,
\eea
(with $\alpha$'s now being treated as functions of $\exb$) and $A_{4}$ is expressed in terms of $A_{1}$ and $A_{2}$ by the relation
\be \label{a4}
A_{4}=\frac{A_{1}}{A_{2}^{2}}-\frac{4\exb}{A_{2}^{2}}\frac{\md A_{2}}{\md\exb}.
\ee
It is immediately obvious from the above expressions that $A_{1}, A_{2}$ and $A_{3}$ are algebraically independent since $\aone, \at$ and $\ath$ are independent functions of $\exb$. Furthermore, since these three $\alpha$-functions are arbitrary, the $A$'s are equally arbitrary functions of $\exb$ (except for $A_{4}$ which is related to $(A_{1}, A_{2})$ by \eqref{a4}) and, therefore, in \eqref{super hamiltonian after canonical transformation} we can forget that $A$'s are described in terms of $\alpha$'s by the above expressions and just treat them as some arbitrary functions of $\exb$. 

The important point is that under a canonical transformation the Poisson brackets retain their form except that the results are expressed in terms of the transformed variables. Since from \eqref{super momentum after canonical transformation} we know that the diffeomorphism constraint retains its classical form even after canonical transformation, the Poisson bracket \eqref{deformed hh bracket}, when written in terms of $(\kxb,\exb)$ acquires the classical form
\be \label{hh bracket after canonical transformation}
\{\tilde{H}[N], \tilde{H}[M]\}=D[\exb(\ep)^{-2}(NM'-N'M)].
\ee
The other two Poisson brackets -- $\{\tilde{H}[N],D[\nx]\}$ and $\{D[\nx],D[M^{x}]\}$ -- continue to have the standard form. Thus, the constraint algebra, which was deformed in terms of the original variables $(\kx, \ex)$, has become classical in terms of $(\kxb, \exb)$. 

Here we would like to refer back to equation \eqref{deformed hh bracket} where we noted that the constraint algebra would be undeformed if $\at\af=1$. From \eqref{super hamiltonian after canonical transformation} we note that after performing the canonical transformation we have obtained precisely this condition (if, for the moment, we forget the distinction between $\alpha$'s and $A$'s, which as noted above is only for book keeping, and think of $\alpha_{i}$ to be the same as $A_{i}$). The conditions \eqref{anomaly freedom one} and \eqref{anomaly freedom two} are automatically taken care of in \eqref{super hamiltonian after canonical transformation} and condition \eqref{a four} (with $\af=\alpha_{6}=1/\at$) is identical to \eqref{a4}.

Thus, instead of bothering with this business of performing canonical transformations, we could have just imposed the condition $\at\af=1$ and would have reached the same conclusion. However, that such would be the case is not a priori obvious and, in fact, does not work in $2+1$ dimensions where, as revealed by some ongoing work, the procedure of canonical tranformation gives more general Hamiltonians compared to imposing a condition similar to $\at\af=1$ outright.

In \cite{canonical transformation} it was further shown that the theory (which, for a deformed constraint algebra, does not have the covariance property of the classical theory under coordinate transformations) in terms of $(\kxb, \exb)$ regains the classical general covariance if the metric corresponding to \eqref{classical metric usual variables} is written not in terms of the original variables $(\ex, \ep)$ as in \eqref{classical metric lqg variables} but in terms of the variables $(\bar{E}^{x}, \ep)$
\be \label{case two metric lqg variables} 
{\rm d}s^2=-N^2{\rm d}t^2+\frac{\ep\,^2}{\bar{E}^{x}}({\rm d}x+N^x{\rm
  d}t)^2+ \bar{E}^{x}{\rm d}\Omega^2.
\ee

We have thus proved that any spherically symmetric theory with a deformed (and anomalous) constraint algebra of the form \eqref{general anomalous algebra} resulting from the Hamiltonian \eqref{generalized hamiltonian lqg variables} can always be brought to a form obeying undeformed constraint algebra by performing a suitable canonical transformation and that the Hamiltonian of the resulting theory has three arbitrary $\ex$ (equivalently, $\exb$) dependent functions. This result is in accord with the findings of \cite{MartinPaily}, where for the deformation factor $\beta=1$ there are three arbitrary functions in the new Hamiltonian (see equations \eqref{deformed algebra}-\eqref{correction functions two} below). The (spherical) symmetry reduced version of general relativity is just one among the three-fold infinity of Hamiltonians corresponding to the choice $A_{1}=A_{2}=A_{3}=1$ (which implies $A_{4}=1$).  

For completeness we now give the equations of motion resulting from the new Hamiltonians. The equations of motion are found using Hamilton's equations $\dot{a}=\{a,H\}$ (in the present case $H\equiv\bar{H}[N]+D[\nx]$ with $\bar{H}[N]$ as given in \eqref{super hamiltonian after canonical transformation} and $D[\nx]$ as given in \eqref{super momentum after canonical transformation}). Evaluating the necessary Poisson brackets we find
\bea \label{equations of motion}
\label{ex dot}
\dot{\bar{E}}^{x} &=& N^{x}\bar{E}^{x'}+2NA_{2}\kp\sqrt{\exb} \\
\label{ephi dot}
\dot{E}^{\varphi} &=& (N^{x}\ep)'+\frac{NA_{1}\kp\ep}{\sqrt{\exb}}+NA_{2}\kxb\sqrt{\exb} \\
\label{kphi dot}
\dot{K}_{\varphi} &=& N^{x}\kp'-\frac{NA_{1}\kp^{2}}{2\sqrt{\exb}}-\frac{NA_{3}}{2\sqrt{\exb}}+\frac{N(\bar{E}^{x'})^{2}}{4A_{2}(\ep)^{2}\sqrt{\exb}}-\frac{NA_{1}(\bar{E}^{x'})^{2}}{8A_{2}^{2}(\ep)^{2}\sqrt{\exb}}+\frac{N'\bar{E}^{x'}\sqrt{\exb}}{2A_{2}(\ep)^{2}} \\
\label{kx dot}
\dot{K}_{x} &=& (N^{x}\kxb)'+\frac{NA_{1}\kp^{2}\ep}{2(\exb)^{3/2}}-\frac{NA_{2}\kp\kxb}{\sqrt{\exb}}+\frac{NA_{3}\ep}{2(\exb)^{3/2}}-\frac{N\ep}{\sqrt{\exb}}\frac{\md A_{3}}{\md\exb}+\frac{N''\sqrt{\exb}}{A_{2}\ep} \nonumber \\
&&-\frac{N'E^{\varphi'}\sqrt{\exb}}{A_{2}(\ep)^{2}}+\left(\frac{N'\bar{E}^{x'}}{\ep\sqrt{\exb}}-\frac{N(\bar{E}^{x'})^{2}}{4\ep(\exb)^{3/2}}+\frac{N\bar{E}^{x''}}{\ep\sqrt{\exb}}-\frac{NE^{\varphi'}\bar{E}^{x'}}{(\ep)^{2}\sqrt{\exb}}\right)\left(\frac{1}{A_{2}}-\frac{A_{1}}{2A_{2}^{2}}\right) \nonumber \\
&&+\frac{N(\bar{E}^{x'})^{2}}{2\ep\sqrt{\exb}}\left(-\frac{1}{A_{2}^{2}}\frac{\md A_{2}}{\md\exb}-\frac{1}{A_{2}^{2}}\frac{\md A_{1}}{\md\exb}+\frac{A_{1}}{A_{2}^{3}}\frac{\md A_{2}}{\md\exb}\right).
\eea

To end this section we would like to make a brief comparison of our result with those of \cite{MartinPaily} where Bojowald et. al., following (and extending) the procedure of \cite{hojman, kuchar geometrodynamics lagrangian}, performed a detailed analysis to construct the most general representation based on the anomaly-free but deformed algebra 
\be \label{deformed algebra}
\{H^{Q}[N],H^{Q}[M]\}=D[\beta(\ex)|\ex|(\ep)^{-2}(NM'-N'M)],
\ee 
(other two Poisson brackets having the standard form). That is, like in \cite{hojman}, one forgets that the deformed algebra \eqref{deformed algebra} can be obtained from a Hamiltonian like \eqref{generalized hamiltonian lqg variables} (with the $\alpha$-functions related by condition \eqref{anomaly freedom one} and \eqref{anomaly freedom two}) and only takes the deformed constraint algebra as the input.  

The conclusion of \cite{MartinPaily} was that the most general representation of the generator of normal deformations or the super-Hamiltonian for the deformed algebra is given by \eqref{generalized hamiltonian lqg variables} with the correction functions $\alpha$ having the following form 
\be \label{correction functions one}
\aone=\sqrt{\beta}c_{1}c_{2}, \quad \at=\sqrt{|\beta|}c_{1},
\ee
\be \label{correction functions two}
\ath=\afo=\text{sgn}(\beta)\frac{\sqrt{|\beta|}}{c_{1}}\left(c_{2}-4\frac{\md~\text{ln}~c_{1}}{\md~\text{ln}~\ex}\right), \quad \af=\alpha_{6}=\text{sgn}(\beta)\frac{\sqrt{|\beta|}}{c_{1}}.
\ee
In the above expression $\beta$, $c_{1}$ and $c_{2}$ are arbitrary functions of $\ex$. The absolute value sign and the $\text{sgn}$ function occur because of the possibility of different orientations of the triads in LQG (a possibility we ignore in the present discussion). Note that the above form of the correction functions is consistent with the requirement for anomaly-free algebra in equations \eqref{anomaly freedom one} and \eqref{anomaly freedom two}. In addition, the $(\kp,\kx)$ independent part of the Hamiltonian had an additional function (denoted $f(\ex)$ in \cite{MartinPaily}) which is the analog of the function $A_{3}$ in \eqref{super hamiltonian after canonical transformation}.

For $\beta=1$, the constraint algebra \eqref{deformed algebra} has the classical structure and the representation would be that of the classical surface deformation algebra. Putting $\beta=1$ in \eqref{correction functions one} and \eqref{correction functions two} we find that the representation so obtained is not the classical representation \eqref{classical hamiltonian constraint lqg variables} (for which all the $\alpha$'s equal unity) but depends on three arbitrary functions $c_{1}(\ex)$, $c_{2}(\ex)$ and $f(\ex)$ just like what we found above. To that extent  our result agrees with that in \cite{MartinPaily}.

However we would like to note that although $\beta=1$ gives new theories with the undeformed (or standard) constraint algebra which are identical to what we have found, in \cite{MartinPaily} it was not realized that even when $\beta(\ex)\neq1$ (so that the constraint algebra is deformed compared to the standard form of \eqref{classical hh algebra}-\eqref{classical dd algebra}), the algebra can be rendered to have the standard form as we have shown here.

Another point worth emphasizing in this context is that although a deformed constraint algebra implies the existence of certain spacetime symmetries because of the fact that the constraint algebra closes (or is non-anomalous), the corresponding symmetry is not the classical diffeomorphism invariance (in other words, solutions of constraints and of equations of motion do not map to other solutions under coordinate transformations for a deformed algebra). What the result of the present work implies is that because a constraint algebra deformed due to the presence of $\ex$ dependent factors can be made to have the standard form, these new theories continue to have classical diffeomorphism invariance (in the $t-r$ plane for the models considered) as a good symmetry.

\section{Proof of inequivalence of the new theories and symmetry reduced general relativity and inclusion of matter}

Although in writing \eqref{super hamiltonian after canonical transformation} we have already given second derivative theories of gravity for spherically symmetric spacetimes which are more general than the spherical reduction of general relativity, there are a few loose ends to be tied. First of all, we have not yet shown that the supposedly new theory is really new. In other words, so far we have not shown that the new and the old theories (by which we mean symmetry reduced general relativity \eqref{classical hamiltonian constraint metric variables}) are not related by a canonical transformation of the form \eqref{freedom in definition of momenta as per hojman et al}. Furthermore, in the last section we focused on the case of spherically symmetric vacuum spacetimes only and we would like to see whether the conclusion of the previous section continues to hold even in the presence of matter.  

We, however, begin by writing down the new theory in terms of the metric variables $(\gxx,\gtt)$ instead of the variables $(\ep,\exb)$. Also, in the previous section we presented only the Hamiltonian of the new theory and we would also like to write down the corresponding Lagrangian. The procedure for obtaining the Lagrangian in terms of the metric variables is straight forward. We use the Hamiltonian \eqref{super hamiltonian after canonical transformation} along with the diffeomorphism constraint \eqref{super momentum after canonical transformation} to obtain the equations of motion for $\exb$ and $\ep$. These are then used to eliminate $\kp$ and $\kxb$ from the Hamiltonian and the diffeomorphism constraint. Performing the inverse Legendre transform one then obtains the Lagrangian in terms of $(\exb, \ep)$ (and their derivatives). Finally, using the correspondence in \eqref{relation between metric variables and lqg variables} (with $\ex$ replaced by $\exb$, as noted just before \eqref{case two metric lqg variables}) this Lagrangian can be written in terms of $(\gtt, \gxx)$.

Since the procedure is straight forward we do not explicitly work through these steps here but directly write down the Lagrangian corresponding to \eqref{super hamiltonian after canonical transformation} 
\bea \label{case two lagrangian metric variables}
\bar{L}&=&\int\md x\bigg[B_{1}\left(-\frac{(\gttd)^{2}\sqrt{\gxx}}{8GN\gtt}+\frac{\nx\gtt'\gttd\sqrt{\gxx}}{4GN\gtt}-\frac{(\nx)^{2}(\gtt')^{2}\sqrt{\gxx}}{8GN\gtt}+\frac{N(\gtt')^{2}}{8G\gtt\sqrt{\gxx}}\right) \nonumber \\
&&+B_{2}\bigg(\frac{\nx\gxx'\gttd}{4GN\sqrt{\gxx}}-\frac{\gttd\gxxd}{4GN\sqrt{\gxx}}+\frac{\nx\gtt'\gxxd}{4GN\sqrt{\gxx}}-\frac{\nx N^{x'}\gtt'\sqrt{\gxx}}{2GN} +\frac{N^{x'}\gttd\sqrt{\gxx}}{2GN} \nonumber \\
&&-\frac{N\gtt''}{2G\sqrt{\gxx}}-\frac{(\nx)^{2}\gtt'\gxx'}{4GN\sqrt{\gxx}}+\frac{N\gtt'\gxx'}{4G(\gxx)^{3/2}}\bigg)+\frac{NB_{3}\sqrt{\gxx}}{2G}-\frac{N(\gtt')^{2}}{2G\sqrt{\gxx}}\frac{\md B_{2}}{\md\gtt}\bigg].
\eea
with $B_{1}=(2A_{2}-A_{1})/A_{2}^{2}$, $B_{2}=1/A_{2}$ and $B_{3}=A_{3}$.

Since $(A_{1},A_{2},A_{3})$ are arbitrary functions of $\exb$ (or $\gtt$ in terms of the metric variables), $(B_{1},B_{2},B_{3})$ are equally arbitrary functions of $\gtt$. It is obvious that the classical Lagrangian \eqref{sph symm lagrangian} is recovered for $B_{1}=B_{2}=B_{3}=1$. Note that for dimensional reasons the $B$'s should be dimensionless functions of $\gtt$. 

We can also write down the Hamiltonian in terms of the usual metric variables. For this we note that from \eqref{case two lagrangian metric variables} the momenta conjugate to the metric variables are 
\bea \label{case two pxx}
\bpxx &=& \frac{B_{2}}{4GN}\frac{(-\gttd+\nx\gtt')}{\sqrt{\gxx}}, \\
\label{case two ptt}
\bptt &=& \frac{1}{4GN}\bigg[B_{1}\left(-\frac{\gttd\sqrt{\gxx}}{\gtt}+\frac{\nx\gtt'\sqrt{\gxx}}{\gtt}\right)+B_{2}\bigg(2N^{x'}\sqrt{\gxx}+\frac{\nx\gxx'}{\sqrt{\gxx}}-\frac{\gxxd}{\sqrt{\gxx}}\bigg)\bigg].
\eea
The phase space variables $(\gxx, \bpxx)$ and $(\gtt, \bptt)$ obey analogues of relations \eqref{classical poisson bracket} and using \eqref{total hamiltonian general expression} we find the (constrained) super-Hamiltonian
\bea \label{case two hamiltonian constraint metric variables}
\bar{H}[N]&=&\int \md x\, N\bigg[-\frac{4G\bpxx\bptt\sqrt{\gxx}}{B_{2}}+\frac{2GB_{1}\gxx^{3/2}(\bpxx)^{2}}{B_{2}^{2}\gtt}-\frac{B_{3}\sqrt{\gxx}}{2G}-\frac{B_{2}\gxx'\gtt'}{4G\gxx^{3/2}}+\frac{B_{2}\gtt''}{2G\sqrt{\gxx}} \nonumber \\
&&-\frac{B_{1}\gtt'^{2}}{8G\sqrt{\gxx}\gtt}+\frac{(\gtt')^{2}}{2G\sqrt{\gxx}}\frac{\md B_{2}}{\md\gtt}\bigg]\approx0, 
\eea
and the super-momentum
\be
\label{case two diffeo constraint metric variables}
\bar{D}[\nx]=\int\md x\,\nx\left[\gtt'\bptt-\gxx'\bpxx-2\gxx\bar{\pi}^{xx'}\right]\approx0.
\ee

Comparing with \eqref{classical hamiltonian constraint metric variables} and \eqref{classical diffeomorphism constraint metric variables} we find that while the diffeomorphism constraint retains its classical form, the Hamiltonian is modified by the presence of $\gtt$-dependent functions. However, the important thing to note is that the super-Hamiltonian \eqref{case two hamiltonian constraint metric variables}, like its classical counterpart \eqref{classical hamiltonian constraint metric variables} is quadratic in canonical momenta and that the spatial derivatives of the three metric are of highest order two. 

Making use of the primary Poisson brackets it is straight forward to verify that the above constraints have the same Poisson bracket algebra as that of the classical theory \eqref{classical hh algebra}-\eqref{classical dd algebra}. This proves that what we have here is a representation of the surface deformation algebra. It remains to be seen whether or not this representation is equivalent to the classical representation given by \eqref{classical hamiltonian constraint metric variables} and \eqref{classical diffeomorphism constraint metric variables}.


According to Hojman et. al. \cite{hojman}, the only freedom in choosing the momentum conjugate to the metric is that given by equation \eqref{freedom in definition of momenta as per hojman et al}. Comparing the momenta in \eqref{case two pxx} and \eqref{case two ptt} with their classical counterparts in equations \eqref{classical momenta pxx} and \eqref{classical momenta ptt} we find that they differ by
\bea \label{difference between classical and quantum momenta case two pxx}
\bpxx-\pxx&=&(B_{2}-1)\left(\frac{-\gttd+\nx\gtt'}{4GN\sqrt{\gxx}}\right), \\
\label{difference between classical and quantum momenta case two ptt}
\bptt-\ptt&=&\frac{1}{4GN}\bigg[(B_{1}-1)\bigg(-\frac{\gttd\sqrt{\gxx}}{\gtt}+\frac{\nx\gtt'\sqrt{\gxx}}{\gtt}\bigg) \nonumber \\
&&+(B_{2}-1)\bigg(2N^{x'}\sqrt{\gxx}+\frac{\nx\gxx'}{\sqrt{\gxx}}-\frac{\gxxd}{\sqrt{\gxx}}\bigg)\bigg]. 
\eea

As per \eqref{freedom in definition of momenta as per hojman et al}, the rhs of the above expressions is to be identified with $\delta\Lambda/\delta\gxx$ and $\delta\Lambda/\delta\gtt$ respectively and, if the representation is unique, $\Lambda\equiv\Lambda[^{3}\mathcal{G}]$, a functional only of the three geometry $^{3}\mathcal{G}$. However, the rhs of both \eqref{difference between classical and quantum momenta case two pxx} and \eqref{difference between classical and quantum momenta case two ptt} explicitly depends on the time derivative of the metric components and these, using \eqref{case two pxx} and \eqref{case two ptt}, can always be expressed in terms of the momentum $\bpxx$ and $\bptt$. 

This explicit dependence of $\Lambda$ (or of the difference between the new and the old momenta) on $(\bpxx, \bptt)$ means that its Poisson bracket with the metric variables $(\gxx,\gtt)$ does not vanish. If the two representations were equivalent (and, thus, related by a canonical transformation) this Poisson bracket would vanish since both the new and the old momenta have identical Poisson brackets with the metric variables. This proves that the representation of the surface deformation algebra as given by the super-Hamiltonian and the super-momentum in \eqref{case two hamiltonian constraint metric variables} and \eqref{case two diffeo constraint metric variables} is inequivalent to the representation obtained by starting with the spherically reduced Einstein-Hilbert Lagrangian and given in \eqref{classical hamiltonian constraint metric variables} and \eqref{classical diffeomorphism constraint metric variables}.

We will again like to emphasize that this inequivalence of the representation is not a contradiction of the theorem in \cite{hojman}. It is merely a reflection of the fact that presence of spherical symmetry in the embedding spacetime and the associated reduction in the number of degrees of freedom (or the size of the phase space) allows more freedom in the construction of generally covariant theories than is possible in the full theory. 

\subsection{Inclusion of matter}
So far we have confined our attention to vacuum spacetimes. However, incorporation of matter is straight forward and will in fact bring in additional $\gtt$ dependent functional degree of freedom. For instance, for spherically symmetric Maxwell field minimally coupled to gravity, the classical (matter) super-Hamiltonian and the super-momentum are 
\bea \label{classical hamiltonian maxwell field}
H_{\mathrm{EM}}[N] &=& \int\md x N\frac{2\pi\sqrt{\gxx}(p^{x})^{2}}{\gtt}, \\
\label{classical diffeo maxwell field}
D_{\mathrm{EM}}[\nx] &=& -4\pi\int\md x\nx A_{x}p^{x'}.
\eea
In the above expressions $p^{x}$ is the momenta canonically conjugate to the radial component of the field $A_{x}$ such that $\{A_{x}(x),p^{x}(y)\}=\delta(x,y)/4\pi$ (other two spatial components $A_{\phi}=A_{\theta}=0$ because of spherical symmetry).

Now it turns out that even the non-classical matter Hamiltonian
\be \label{non classical hamiltonian maxwell field} 
\bar{H}_{\mathrm{EM}}[N]=\int\md x N\frac{2\pi B_{4}\sqrt{\gxx}(p^{x})^{2}}{\gtt},
\ee
(where $B_{4}$ is yet another arbitrary function of $\gtt$) when appended to the non-classical gravitational Hamiltonian \eqref{case two hamiltonian constraint metric variables} satisfies the surface deformation algebra of equations \eqref{surface deformation algebra hh}-\eqref{surface deformation algebra dd} showing that non-uniqueness of representation is not limited to vacuum spacetimes (of course, we also add the matter super-momentum \eqref{classical diffeo maxwell field} to the gravitational super-momentum \eqref{case two diffeo constraint metric variables}). 

It should be clear that this will hold for other matter degrees of freedom as well. As a second example consider the inclusion of a scalar field $\Phi$. The super-Hamiltonian and the super-momentum for the scalar field are given by:
\bea \label{scalar field hamiltonian}
H_{\mathrm{s}}[N] &=& 4\pi\int\md x\left[\frac{\pi_{\Phi}^{2}}{22\gtt\sqrt{\gxx}}+\frac{\gtt\Phi{'2}}{2\sqrt{\gxx}}+\frac{\gtt\sqrt{\gxx}V(\Phi)}{2}\right], \\
\label{scalar field diffeo}
D_{\mathrm{s}}[\nx] &=& 4\pi\int\md x\nx\pi_{\Phi}\Phi'.
\eea
In the above equations, $\pi_{\Phi}$ is the momentum conjugate to $\Phi$ and $V(\Phi)$ is the potential. 

Even for scalar field it turns out that the following super-Hamiltonian
\be \label{modified scalar field hamiltonian}
\bar{H}_{\mathrm{s}}[N]=4\pi\int\md x\left[\frac{B_{5}\pi_{\Phi}^{2}}{22\gtt\sqrt{\gxx}}+\frac{\gtt\Phi{'2}}{2B_{5}\sqrt{\gxx}}+\frac{B_{6}\gtt\sqrt{\gxx}V(\Phi)}{2}\right],
\ee
when appended to the new gravitational super-Hamiltonian \eqref{case two hamiltonian constraint metric variables} (and, correspondingly, the scalar field super-momentum is combined with the gravitational super-momentum \eqref{case two diffeo constraint metric variables}) continues to satisfy the surface deformation algebra \eqref{classical hh algebra}-\eqref{classical dd algebra}. Here $B_{5}\equiv B_{5}(\gtt)$ and $B_{6}\equiv B_{6}(\gtt)$ are arbitrary functions of $\gtt$. In line with the comment below equation \eqref{generalized hamiltonian lqg variables}, we leave the scalar super-momentum unmodified. 

As for the gravitational sector, one can say that the reduced phase space resulting from the imposition of spherical symmetry gives more freedom in the construction of possible generally covariant matter theories. And in consonance with the result in \cite{hojman} this freedom disappears when no symmetry is imposed on the spacetime. From the point of view of the Poisson brackets it is easy to see why the modified matter Hamiltonians - \eqref{non classical hamiltonian maxwell field} for the Maxwell field and \eqref{modified scalar field hamiltonian} for the scalar field continue to satisfy the surface deformation algebra. 

First we note that even after inclusion of the modifications, the matter field continues to be minimally coupled to gravity (by which we mean there is no derivative coupling involved). The other thing to note is that all the modification functions $\alpha$'s (or the $A$'s and $B$'s) are scalar quantities, being defined in terms of $\gtt$ (or $\ex$) which has density weight zero ($\ep=\sqrt{\gtt\gxx}$ has density weight one). As already remarked, the super-momentum is left unmodified and only the super-Hamiltonian is modified. Also, the gravitational part of the constraints are independent of the matter degrees of freedom while in the matter sector, only the matter Hamiltonian depends on the gravitational degrees of freedom (the three metric). Next we write the total (modified) super-Hamiltonian as the sum of gravitational and matter parts $\bar{H}[N]=\bar{H}_{\mathrm{G}}[N]+\bar{H}_{\mathrm{M}}[N]$ and similarly $D[\nx]=D_{\mathrm{G}}[\nx]+D_{\mathrm{M}}[\nx]$.

When evaluating the $H-H$ Poisson bracket we find that because of the non-derivative nature of the coupling, the part $\{\bar{H}_{\mathrm{G}}[N],\bar{H}_{\mathrm{M}}[M]\}+\{\bar{H}_{\mathrm{M}}[N],\bar{H}_{\mathrm{G}}[M]\}=0$. We have already evaluated the bracket between two gravitational Hamiltonians $\{H_{\mathrm{G}}[N],H_{\mathrm{G}}[M]\}$ and it turns out that even with modification the bracket between two matter Hamiltonians evaluates to the corresponding matter diffeomorphism in the required form. Similarly, for the $D-H$ bracket one can convince oneself  that the modifications do not affect the structure of the bracket and the question of a possible modification of the structure of the $D-D$ bracket does not arise as the diffeomorphism constraint is left unmodified.

\section{Motivation for the new theories: Inverse triad corrections in LQG}

Although we have already given the (mathematical) derivation of the new theories in the previous two sections, in this section we will give the basic physics motivation originating in LQG which led to the identification of new generally covariant theories for spherically symmetric spacetimes. As already indicated there, our choice of the variables was motivated from the studies in LQG where the classical theory is first recast in terms of the $\mathfrak{su}(2)$ Ashtekar variables. For spherically symmetric spacetimes these are the components of the densitized triad $(\ep, \ex)$ and their conjugate variables $(\kp, \kx)$ which are related to the extrinsic curvature components $K_{ab}$ (details on the LQG formulation of spherically symmetric spacetimes can be found in \cite{SphSymmstates, SphSymmHam}). 

Most of the expressions of interest to us have already appeared in section 4 and, therefore, we will not repeat them all here. For instance, the relation of the densitized triad variables $(\ep, \ex)$ to the usual metric variables is given in \eqref{relation between metric variables and lqg variables} whereas the diffeomorphism constraint is given in \eqref{diffeomorphism constraint lqg variables}. Because of its importance in the discussion to follow, we reproduce the Hamiltonian constraint:
\bea \label{classical hamiltonian constraint lqg variables2}
H[N] &=& -\frac{1}{2G}\int \md x\, N\bigg[\frac{\kp^2\ep}{\sqrt{\ex}}+2\kp\kx\sqrt{\ex}+\frac{\ep}{\sqrt{\ex}} -\frac{(E^{x'})^{2}}{4\ep\sqrt{\ex}}-\frac{E^{x''}\sqrt{\ex}}{\ep} \nonumber \\
&&+\frac{E^{\varphi'}E^{x'}\sqrt{\ex}}{(\ep)^{2}} \bigg]\approx0,
\eea

As we already alluded to in section 4, in LQG the presence of inverse components of the triad variable $\ex$ in the Hamiltonian constraint \eqref{classical hamiltonian constraint lqg variables2} leads to certain corrections in the Hamiltonian (known as the inverse triad corrections) since the operator version of $\ex$ has discrete spectrum containing zero and, therefore, cannot be inverted trivially. To be more specific, an orthonormal basis for spherically symmetric spacetimes in the connection representation used in LQG is given by \cite{SphSymmstates}
\begin{equation} \label{GaugeInvSpinNetwork}
 T_{g,k,\mu}=\prod_{e\in g} \exp\left({\textstyle\frac{1}{2}}i k_e
\smallint_e(A_x+\eta')\md x\right)  \prod_{v\in g}
\exp(i\mu_v \gamma K_{\vp}(v))
\end{equation}
with $k_e\in\mathbb{Z}$ and positive real labels $\mu_v$ on edges
$e$ and vertices $v$, respectively. The action of the operator valued triad component $(\hat{\ex},\hat{\ep})$ on these states is 
\bea \label{ex spectrum}
\hat{\ex}(x) T_{g,k,\mu} &=& \gamma\lp^2
\frac{k_{e^+(x)}+k_{e^-(x)}}{2} T_{g,k,\mu}, \\
\label{ep spectrum}
\int_{\cal I}\hat{E}^{\vp}T_{g,k,\mu} &=& \gamma\lp^2
\sum_{v\in{\cal I}} \mu_v T_{g,k,\mu}
\eea
where $\ell_{\rm P}^2=G\hbar$ is the Planck length squared and $\gamma$ is the Barbero-Immirzi parameter while
$e^{\pm}(x)$ denote the neighboring edges to a point $x$ distinguished from each other using a given orientation of the radial line. 

As seen from \eqref{ex spectrum}, the spectrum of the operator $\hat{\ex}$ contains zero. This implies that in the quantum theory there is no direct quantization of the inverse of operator $\hat{\ex}$. Following the methods available in the full theory \cite{ThiemannQSD} we can, nevertheless, construct suitable operator version of $(\hat{\ex})^{-1}$ which reproduces the inverse of the triad variable $(\ex)^{-1}$ in the classical limit. 

The key observation is that classically we can write 
\be \label{inverse ex in terms of volume and ax}
4\pi\gamma G{\rm
sgn}(E^x)E^{\vp}/\sqrt{|E^x|}= \{A_x,V\},
\ee where $V=4\pi\int\md x \sqrt{|E^x|}E^{\vp}$ is the classical expression for
volume in spherically symmetric setting and where $A_{x}=\gamma\kx-\eta'$ is the connection component conjugate to $\ex$. (In the original formulation, apart from the Hamiltonian and the diffeomorphism constraints there is an additional Gauss constraint in the theory resulting from the use of Ashtekar variables and, correspondingly, there is an extra pair of conjugate variables $(\eta,P^{\eta})$. If we solve the Gauss constraint classically, then this pair gets eliminated and one is left with $\kx$ as the variable conjugate to $\ex$. Going into more details would be an unnecessary digression and the interested reader can refer to \cite{SphSymmstates, SphSymmHam} for more details.)

In the quantum theory one can get a handle on the inverse of operator $\hat{\ex}$ by `quantizing' the right side of \eqref{inverse ex in terms of volume and ax}. The result is (see \cite{LTB1} for details)
\be \label{inverse ex operator}
 \widehat{\int_{\cal I} \frac{E^{\varphi}{\rm sgn}(E^x)}{\sqrt{|E^x|}}} =
\frac{-i}{2\pi\gamma G\hbar} {\rm
tr}(\tau_3h_x[h_x^{-1},\hat{V}]),
\ee
where $h_{x}=\mathrm{exp}(\tau_{3}A_{x})$ is the holonomy of $A_{x}$ and $\tau_{3}=-i\sigma_{3}/2$ ($\sigma_{3}$  being the $z$-component of the Pauli matrices). The corresponding eigenvalues of this operator are
\begin{equation}
 \left(\widehat{\int_{\cal I}
 \frac{E^{\varphi}{\rm sgn}(E^x)}{\sqrt{|E^x|}}}\right)_{k,\mu} =
 2\sqrt{\gamma}\ell_{\rm P}
 |\mu_v|\left(\sqrt{|k_{e^+(v)}+k_{e^-(v)}+1|}-
\sqrt{|k_{e^+(v)}+k_{e^-(v)}-1|}\right).
\end{equation}

On comparison with \eqref{ex spectrum} and \eqref{ep spectrum} we see that, in a semi-classical approach, we can parameterize the inverse of the operator $\hat{\ex}$ in terms of a correction function $\alpha(\ex)$ such that 
\begin{equation} \label{alpha}
\alpha({E^{x}}):=
\left(\widehat{\frac{1}{\sqrt{|E^x|}}}\right)_{k(E^x)}
\left(\sqrt{|\hat{E}^x|}\right)_{k(E^x)} =
2\frac{\sqrt{|E^{x}+\gamma \lp^{2}/2|}-
\sqrt{|E^{x}-\gamma \lp^{2}/2|}}{\gamma \lp^{2}}\sqrt{|E^{x}|}
\end{equation}
Note that $\alpha(\ex)\rightarrow1$ for $\gamma\lp^{2}\ll\ex$. The above derivation of the inverse triad correction  shows that, unlike what is sometimes thought, these effects are \emph{not} put in an adhoc manner even for symmetry reduced models.

After this brief digression on the explicit form of $\alpha(\ex)$, we now get to the main point of the discussion. At a semi-classical level the LQG effects arising due to the presence of inverse components of $\hat{\ex}$ operator are incorporated in the Hamiltonian by making the replacement $1/\ex\rightarrow\alpha(\ex)/\ex$. This has the correct classical limit and incorporates certain quantum gravity effects as well. To keep things general, in the Hamiltonian, occurences of $\ex$ with different powers are corrected with different $\alpha$'s, $1/(\ex)^{i}\rightarrow\alpha_{i}/(\ex)^{i}$ (where different $\alpha_{i}$ have the same general form of \eqref{alpha} but can differ from each other through quantization ambiguities \cite{MartinQuantAmb}) and the Hamiltonian constraint incorporating inverse triad corrections takes the form:
\bea \label{inverse triad corrected hamiltonian}
H[N] &=& -\frac{1}{2G}\int \md x\, N\bigg[\frac{\alpha\kp^2\ep}{\sqrt{\ex}}+2\bar{\alpha}\kp\kx\sqrt{\ex}+\frac{\alpha\ep}{\sqrt{\ex}} -\frac{\alpha(E^{x'})^{2}}{4\ep\sqrt{\ex}}-\frac{\bar{\alpha}E^{x''}\sqrt{\ex}}{\ep} \nonumber \\
&&+\frac{\bar{\alpha}E^{\varphi'}E^{x'}\sqrt{\ex}}{(\ep)^{2}} \bigg]\approx0, \\
\eea
In LQG, the action of the diffeomorphism constraint is directly represented on the quantum states through group averaging and therefore the diffeomorphism constraint is left unmodified. 

This gives the physical motivation behind the kind of corrections that were considered in section 4 (of course the Hamiltonian we considered there was more general than \eqref{inverse triad corrected hamiltonian} since we wanted to construct the most general theory with such corrections). These corrections have been investigated in several works from a semi-classical point of view \cite{LTB1, LTB2, modifiedhorizon, einstein maxwell}. As should be clear by now, the effect of these corrections in the Hamiltonian is that, in general, the constraint algebra is deformed (and possibly anomalous) by the presence of $\ex$ dependent functions in the $H-H$ Poisson bracket (see \eqref{general anomalous algebra}) \cite{LTB2, modifiedhorizon}. Equating the anomalous part to zero leads to certain conditions which when used in the $D-H$ bracket implies that it retains the classical form of \eqref{classical dh algebra} (the $D-D$ bracket obviously remains unchanged since the diffeomorphism constraint is left unmodified). 

In these early works it was not known that the deformed constraint algebra can be made to have the standard form of equations \eqref{classical hh algebra}-\eqref{classical dd algebra} and the analysis of the equations of motion and their solutions suggested that the spacetime structure is modified because of the deformed algebra \cite{modifiedhorizon}. However, as suggested in \cite{canonical transformation} (for a special case) and as explicitly generalized to the case of independent $\alpha_{i}$'s in section 4, the $\ex$ dependent deformation can always be gotten rid of by a suitable canonical tranformation of the geometrodynamical phase space and the canonically transformed theory has the standard properties under spacetime diffeomorphisms once the metric is written using the transformed variables $\exb$ instead of the original variable $\ex$ (see \eqref{case two metric lqg variables}).

It should, however, be noted that inverse triad corrections correspond to only one kind of quantum gravity correction. More generally, one will also need to worry about the fact that in LQG connection components do not have a direct representation as operators on the Hilbert space but only their exponentials (holonomies) are well defined operators. These \emph{holonomy effects} should also be included in the Hamiltonian and inclusion of these effects also leads to deformed constraint algebra \cite{JuanThesis}. And, in general, it seems that these connection/curvature dependent deformation factors cannot be taken care of completely (by the procedure of canonical transformations) without introducing some other complication in the theory (in \cite{canonical transformation} it was found that, unlike the classical Hamiltonian, this leads to the appearence of derivative of the momentum component in the Hamiltonian). This suggests that when all the different kinds of quantum corrections are taken into account, the spacetime structure will indeed get modified.

Limiting to the case of inverse triad corrections, we can still ask at what scale are these effects expected to be important? A look back at \eqref{alpha} will suggest that these effects will be dominant when $\ex\approx\gamma\lp^{2}$ since $\alpha\rightarrow1$ very quickly once $\ex>\gamma\lp^{2}$. However, this conclusion is a bit premature since it ignores the underlying discreteness of the full theory (one of the main features of LQG). One expects that since the spacetime is smooth at large distance scales, the underlying discreteness should get refined as one moves to macroscopic scales. 

Symmetry reduced models, by their very construction are blind to these refinements along the symmetry direction and their effect can be taken into account by using the so-called lattice refinement methods which have been studied quite a bit in LQC \cite{MartinLatticeRefine, MartinInhomogeneities} and also to some extent in spherically symmetric models \cite{modifiedhorizon, einstein maxwell}. The point is that if we imagine a macroscopic orbit of size $|\ex|$ to be made up of $\mathcal{N}(\ex)$ underlying discrete plaquettes then the main effect of these schemes is that $\ex$ in \eqref{alpha} gets replaced by $\ex/\mathcal{N}(\ex)$ (the plaquette size). In such a case $\ex/\mathcal{N}(\ex)\approx\gamma\lp^{2}$ is the scale where quantum gravity effects are important and since $\mathcal{N}(\ex)$ can be large, the corresponding effects can be felt at scales much greater than Planck length (the exact scale depending on the refinement scheme).

\section{Significance of the new models and going beyond spherical symmetry}

The first question to ask whenever presented with new models/theories is what is their significance? The most obvious significance of the models is that when interpreted in terms of LQG corrections like the inverse triad corrections so that the arbitrary functions $\alpha_{i}$'s are determined by the expression \eqref{alpha}, these models can give an understanding of the role and the importance of these corrections. The Hamiltonian \eqref{super hamiltonian after canonical transformation} can be thought of as an effective Hamiltonian which incorporates certain LQG corrections.

More importantly, since the constraint algebra even after incorporating LQG corrections has the standard form \eqref{classical hh algebra}-\eqref{classical dd algebra}, it implies that the quantum corrected theory retains the underlying diffeomorphism covariance under coordinate transformations in the $t-r$ plane (which was not completely the case in earlier investigations in \cite{LTB1,LTB2,MartinPaily,JuanThesis, einstein maxwell}). This is in accordance with what is generally expected, that even if in the deep quantum gravity regime the usual notions of differential geometry do not survive, the more algebraic notion of diffeomorphism invariance as encoded in the constraint algebra should survive. 

From a generic point of view, the question of significance of these theories can be answered at two levels: 1) what use these models can be put to? and 2) whether it is possible to go beyond spherical symmetry? The significance of having more than one model/theory at ones disposal can hardly be over emphasized. Presence of more than one models within a given framework (framework of diffeomorphism invariance in the present case) gives an opportunity to explore the framework across these models and thus helps to build intuition.

Since general relativity, by construction, is such a tight framework there is not much scope to have this freedom and to go beyond general relativity while retaining general covariance, one usually needs to include higher derivative/curvature terms in the action or to build theories like scalar-tensor theories. Although these constructions are well motivated, it is still desirable to have more theories without incorporating higher derivative terms (which, in general, make computations much more complex) or which do not require the need to include new matter degrees of freedom. In other words, ideally one would like to be as close to the original theory as possible but still have some freedom available. Within the context of spherically symmetric spacetimes at least, we now have three-fold infinity of theories in vacuum (with additional freedom in the presence of matter) which are generally covariant (in the $t-r$ plane) and invlove no more than second derivatives of the metric.

As is well known, the spherical symmetry ansatz has played quite an important role in the understanding of general relativity. Given the complicated nature of Einstein's equations, these models provide the rare scenario where these equations can be solved explicitly. With more than one second derivative theories available for spherically symmetric spacetimes, it can be tested how general the conclusions drawn in spherically symmetric general relativity are. By suitably choosing the functions $(A_{1}(\exb),A_{2}(\exb),A_{3}(\exb))$ in \eqref{super hamiltonian after canonical transformation} (correspondingly $(B_{1}(\gtt),B_{2}(\gtt),B_{3}(\gtt))$ in \eqref{case two hamiltonian constraint metric variables} one can construct a variety of solutions (not present in spherically reduced general relativity). One can explore how consistent are the conclusions drawn from these solutions compared to the solutions in spherically symmetric general relativity.

Apart from exploring the classical aspects of these theories for different choices of $A_{1}$, $A_{2}$ and $A_{3}$, one can also explore the quantum gravity aspects of these theories. As is well known, a complete quantization of general relativity is still a distant goal. As a result a lot of effort is expended on understanding aspects of quantum gravity from symmetry reduced models - the, so called, mini-superspace (homogeneous) cosmological models with finitely many degrees of freedom and the midi-superspace models which, despite symmetry reduction, have infinit degrees of freedom (spherically symmetric models being one example). 

However, as mentioned before, if one insists on working within the context of Einstein's theory then one is stuck with a single spherically symmetric Hamiltonian to quantize. Based on the study of just one Hamiltonian it is very difficult to conclude about the robustness of the results. Availability of more than one Hamiltonians is therefore a highly desired property. Perturbative quantum field theory (QFT) providing a good case in point where investigations of numerous models, even when not all of these describe nature, has helped to build intuition about QFTs which led to the subsequent development of the field.

As it is, even the spherically symmetric general relativity is not fully under control when it comes to quantization. Using the freedom in $(A_{1},A_{2},A_{3})$, one can make a suitable choice for these functions such that the resulting Hamiltonian is simpler compared to the Hamiltonian of spherically symmetric general relativity. It can then be hoped that the simplified Hamiltonian will be easier to quantize.

However, despite all the above mentioned usefulness of the new theories of this paper there remains the caveat that these new theories require spherically symmetric spacetimes. This, in a certain sense, limits the utility of these models since (generally covariant) spherically symmetric models do not allow true dynamical degrees of freedom. Ideally one would like to go beyond spherical symmetry and see if such freedom remains. From the result of \cite{hojman} it is obvious that such a freedom does not exist for a generic spacetime. 

However, one possible way to go beyond spherical symmetry might be to consider spacetimes which have lesser degree of symmetry (less number of Killing vectors), for instance, spacetimes which allow rotations and try to see if the analysis similar to that of \cite{hojman} allows for more general theories even there. Another direction could be to explore the possibility of introducing non-spherical linear perturbations. In the presence of the $\alpha(\ex)$ modifications, the constraint algebra, as before, would be deformed. However, now the important question would be whether, in the presence of linear (non-spherical) perturbations, the constraint algebra can be straightened out (as was the case for perfect spherical symmetry). This, as should be obvious, will be quite a difficult task. However, if such a generalization turns out to be possible even at linear order in perturbations and possibly with reduced freedom for arbitrary functions, it would still be of great value. We hope to explore these issues in the future.

\section{Conclusions}

According to the work of Hojman et. al \cite{hojman}, there is not much freedom as far as the \emph{canonical} form that a generally covariant theory obeying the surface deformation algebra in equations \eqref{surface deformation algebra hh}-\eqref{surface deformation algebra dd} can take. However, we have shown that for symmetry reduced models, specifically, for spherically symmetric spacetimes, this is not the case and that considerable freedom is available in constructing generally covariant theories. We found that there is a three-fold infinity of freedom available in these theories as revealed by the presence of three arbitrary $\gtt$ dependent functions in the Hamiltonian. 

It should be noted that this is \emph{not} in contradiction with the result in \cite{hojman} where the proof is for full general relativity with its $12\infty^{3}$
dimensional phase space (before imposing constraints) whereas for the (spherical) symmetry
reduced sector of this paper the phase space is much smaller (only $4\infty$ dimensional
before imposing constraints). The uniqueness result of Hojman et. al. therefore need not apply. As should be obvious, for the full theory with no Killing vectors the result in \cite{hojman} will continue to hold.

Following and generalizing the procedure of \cite{canonical transformation} we also showed that a theory with deformed constraint algebra (with $\gtt$ dependent deformation factors) could always be made to obey the standard constraint algebra by absorbing the deformation factor through a suitable canonical transformation. Interestingly, the super-momentum (or the diffeomorphism constraint) retained its classical form even after the canonical transformation. Most interestingly, the new class of super-Hamiltonians continue to have the basic features of the super-Hamiltonian based on general relativity (and the ones used by Hojman et. al. in their proof) - these are quadratic in canonical momenta and are quadratic in spatial derivatives of the three-metric.

In LQG, where these other representations were first obtained, concerns have occasionally been raised as to whether the different versions of the Hamiltonian, giving the classical constraint algebra are really different or are they equivalent to the classical Hamiltonian. We proved that the new representation is inequivalent to the classical representation by showing that the momenta in the new theory differ from the momenta in the old theory by terms which involve time derivatives of the three metric (see equations \eqref{difference between classical and quantum momenta case two pxx} and \eqref{difference between classical and quantum momenta case two ptt}. We also showed that the results continue to hold even in the presence of Maxwell field and the scalar field (obeying spherical symmetry).  

Since the results presented here depend crucially on the spacetime being spherically symmetric, it seems fair to suppose that the existence of the Killing vectors might be playing a crucial role in the construction of the inequivalent theories. However, at this stage it is not clear as to the exact role played by the Killing vectors. To understand this, it might be useful to investigate the question as to how much freedom is available for other models where the number of Killing vectors is different? For instance, one could consider models with weaker symmetry -- spacetimes with rotation, where the number of Killing vectors is less.

Ideally one would also like to understand the exact geometric character of the generalized Lagrangian presented in \eqref{case two lagrangian metric variables}. Unfortunately, so far we have been unable to do so. This should not be very surprising considering the fact that the modifications involve only one component $\gtt$ of the metric tensor.

Having demonstrated the existence of infinitely many generally covariant second derivative theories for spherically symmetric spacetimes, one would like to know the possible implications of such a result. One could raise the objection that there is no fundamental theory that is applicable only to spherically symmetric spacetimes, and therefore the present work is just a mathematical curiosity. That such is not the case is clear from a  consideration of symmetry reduced toy models like the homogeneous and isotropic FRW cosmology which describes the background cosmology. Similarly, a lot of interesting results and conclusions in classical general relativity and black hole physics are based on investigations of spherically symmetric models. The alternative Lagrangians/Hamiltonians of the present work can act as very useful toy models in which to test whether the conclusions so obtained are specific to general relativity or apply to diffeomorphism invariant theories in general.  

From another point of view, the problem of quantum gravity is still unsolved and one has to make frequent use of toy models in which to apply ideas of quantum gravity. However, in general, one is either stuck with general relativity as the only theory in which to construct such toy models (which are not many) or to consider higher derivative theories of gravity. The constructions of this work thus provide a unique possibility of working with models which are close enough to general relativistic models (in that they do not involve higher derivatives) and are yet numerous enough because of the presence of arbitrary functions. The implications of midi-superspace quantization can now be tested across various models as one is not limited to the unique midi-superspace model based on general relativity. In addition, by making a suitable choice for the arbitrary functions $A_{1}$, $A_{2}$ and $A_{3}$ in \eqref{super hamiltonian after canonical transformation} (or for $B_{1}$ $B_{2}$ and $B_{3}$ in \eqref{case two hamiltonian constraint metric variables}) the resulting Hamiltonian can be simplified, which could then aid in its quantization.

\section*{Acknowledgements}
The author would like to thank A.P. Balachandran and Martin Bojowald for useful discussions and for there comments on the manuscript. This work is supported under DST-Max Planck India Partner Group in Gravity and Cosmology.


\begin{thebibliography}{99}

\bibitem{hojman}
S.~A.~Hojman, K.~Kuchar and C.~Teitelboim,
Geometrodynamics Regained,
  Annals Phys.\  {\bf 96}, 88 (1976).

\bibitem{lovelock}
D.~Lovelock
The Einstein Tensor and Its Generalizations,
Jour.\ Math.\ Phys.\ {\bf 12}, 498 (1971).

\bibitem{modifiedhorizon}
M.~Bojowald, G.~M.~Paily, J.~D.~Reyes and R.~Tibrewala,
  Black-hole horizons in modified space-time structures arising from canonical quantum gravity,
  Class.\ Quant.\ Grav.\  {\bf 28}, 185006 (2011) (arXiv:1105.1340).

\bibitem{MartinPaily}
M.~Bojowald and G.~M.~Paily,
 Deformed General Relativity and Effective Actions from Loop Quantum Gravity,
  Phys.\ Rev.\ D {\bf 86}, 104018 (2012) (arXiv:1112.1899).

\bibitem{canonical transformation}
R.~Tibrewala,
  Inhomogeneities, loop quantum gravity corrections, constraint algebra and general covariance,
  Class.\ Quant.\ Grav.\  {\bf 31}, 055010 (2014) (arXiv:1311.1297).

\bibitem{adm}
R.~Arnowitt, S.~Deser and C.~Misner, The dynamics of general relativity,
Gen.\ Rel.\ Grav.\ {\bf 40}, 1997 (2008) (arXiv:gr-qc/0405109).

\bibitem{kuchar geometrodynamics lagrangian}
K.~Kuchar,
Geometrodynamics regained - a lagrangian approach,
  J.\ Math.\ Phys.\  {\bf 15}, 708 (1974).

\bibitem{SphSymmstates}
M.~Bojowald,
  Spherically symmetric quantum geometry: States and basic operators,
  Class.\ Quant.\ Grav.\  {\bf 21}, 3733 (2004) (arXiv:gr-qc/0407017).

\bibitem{SphSymmHam}
M.~Bojowald and R.~Swiderski,
  Spherically symmetric quantum geometry: Hamiltonian constraint,
  Class.\ Quant.\ Grav.\  {\bf 23}, 2129 (2006) (arXiv:gr-qc/0511108).

\bibitem{ThiemannQSD}
T.~Thiemann, Quantum spin dynamics (QSD), Class. Quantum Grav. {\bf 15}, 839 (1998) (arXiv:gr-qc/9606089).

\bibitem{LTB1}
M.~Bojowald, T.~Harada and R.~Tibrewala,
  Lemaitre-Tolman-Bondi collapse from the perspective of loop quantum gravity,
  Phys.\ Rev.\ D {\bf 78}, 064057 (2008) (arXiv:0806.2593).

\bibitem{MartinQuantAmb}
M.~Bojowald, Quantization ambiguities in isotropic quantum geometry, Class. Quantum Grav. {\bf 19} 5113 (2002)
(arXiv:gr-qc/0206053).

\bibitem{LTB2}
M.~Bojowald, J.~D.~Reyes and R.~Tibrewala,
  Non-marginal LTB-like models with inverse triad corrections from loop quantum gravity,
  Phys.\ Rev.\ D {\bf 80}, 084002 (2009) (arXiv:0906.4767).

\bibitem{einstein maxwell}
R.~Tibrewala,
  Spherically symmetric Einstein-Maxwell theory and loop quantum gravity corrections,
  Class.\ Quant.\ Grav.\  {\bf 29}, 235012 (2012) (arXiv:1207.2585). 

\bibitem{JuanThesis}
J.~D.~Reyes, Spherically symmetric loop quantum gravity: connections to 2-dimensional models and
applications to gravitational collapse, PhD Thesis The Pennsylvania State University, University Park, PA,
USA (2009). 

\bibitem{MartinLatticeRefine}
M.~Bojowald, D.~Cartin and G.~Khanna, Lattice refining loop quantum cosmology, anisotropic models and
stability, Phys. Rev. D {\bf 76} 064018 (2007) (arXiv:0704.1137).

\bibitem{MartinInhomogeneities}
M.~Bojowald, Loop quantum cosmology and inhomogeneities, Gen. Rel. Grav. {\bf 38} 1771 (2006)
(arXiv:gr-qc/0609034).



\end{thebibliography}
\end{document}